\documentclass[useAMS,usenatbib]{mn2e}

\def\f{\varphi}

\def\g{\gamma}

\usepackage{graphicx}
\title[Magnetocentrifugal acceleration in pulsar magnetosphere]{Magnetocentrifugal acceleration of  bulk motion of plasma in pulsar magnetosphere}
\author[S. V. Bogovalov]{S.V.Bogovalov\thanks{E-mail:
svbogovalov@mephi.ru}\\
National Research Nuclear University (MEPHI), Moscow, Russia}

\begin{document}

\date{}

\pagerange{\pageref{firstpage}--\pageref{lastpage}} \pubyear{2014}

\maketitle

\label{firstpage}

\begin{abstract}

Acceleration of bulk motion of plasma due to magneto centrifugal mechanism is investigated for 
different shapes of the field lines of 3-D magnetic field. It is shown that  this 
mechanism can be efficient provided that the field line is twisted into direction of rotation. According to the last results of the numerical modelling a fraction  of the field lines with such geometrical shape apparently exist in the pulsar magnetosphere close to the last closed field line. In this case the Lorentz factor $\gamma$ increases with the radius $r$ along these field lines as  $\gamma\sim (1-(r\Omega/c)^2)^{-1}$, where $\Omega$ - angular velocity of the pulsar rotation. The magneto centrifugal mechanism provides acceleration of the electrons to the energy  which is sufficient to explain the observed VHE pulsed $\gamma$-rays from  Crab pulsar. 

%Equation defining Lorentz factor of plasma is obtained for the case of plasma motion along rotating field lines of the magnetic field for an 
%arbitrary  shape of the lines.
%Frozen in condition of the magnetic field into the plasma is assumed. Acceleration of plasma due to magneto centrifugal force is investigated for 
%different shapes of the field lines. It is shown that  the Lorentz factor $\gamma$ increases with the radius
%$r$ as  $\gamma\sim (1-(r\Omega/c)^2)^{-1}$, where $\Omega$ - angular velocity of the pulsar rotation provided
%that the field line is bent into direction of rotation. A fraction  of the field lines with such a shape always exist in the pulsar magnetosphere. 
%It is shown that the magneto centrifugal acceleration provides acceleration of the particles at least to the energy equal to... which is sufficient to explain the observed VHE pulsed radiation from  Crab pulsar. 
\end{abstract}

\begin{keywords}
pulsars, plasma acceleration, magnetohydrodynamics, centrifugal acceleration
\end{keywords}

\section{Introduction}
Discovery of pulsed  $\g$ -rays with energy $\sim 400 ~\rm GeV$ from the Crab pulsar by Cherenkov telescopes MAGIC and VERITAS \citep{magic,veritas} initiated search of mechanisms  providing acceleration of electrons and generation of radiation of such energy \citep{aharonian,bednarek}. Nevertheless, no widely accepted mechanism of the particle acceleration has been  found up to now \citep{hirotani}. In this paper we consider the well known mechanism of magneto centrifugal acceleration of plasma.
This mechanism has an evident mechanical analogy \citep{blandford}. Every electron in the magnetic field can be considered as a bear on a wire.  The bear slides along the wire and is centrifugally accelerated at the rotation of the wire. However, this mechanism looks not efficient for acceleration of relativistic plasma  in the pulsar magnetosphere. Analysis shows that for the relativistic plasma with high magnetization when the ratio of the Poynting flux over the density of the kinetic energy flux of the plasma $\sigma \gg 1$ but $\sigma/\g_0^2 \ll 1$ the centrifugal mechanism  is disappointingly inefficient. The magnetic field line appears twisted by the inertia of the plasma and electromagnetic field  into Archimedean spiral which does not provide efficient interaction between plasma and the magnetic field. 
\begin{figure} 
\begin{center}
\includegraphics[width=84mm]{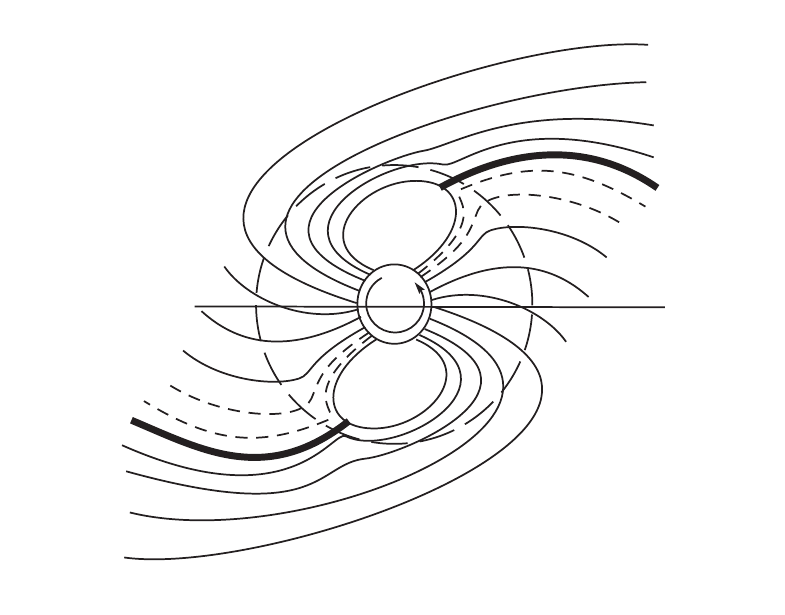}
 \caption{Schematic structure of the magnetic field of an oblique rotator in the equatorial plane. View along the axis of rotation. The field lines with positive twist are shown as dashed ones. Thick lines
show the field lines (current sheets) separating the open field lines of opposite polarities. }
 \label{fig0}
\end{center}
\end{figure} 

In our earlier work \citep{bog2001} we argued that  the inefficiency of the centrifugal acceleration is proven only for the axisymmetric models. Real pulsar magnetospheres are not axisymmetric. There are field lines in a dipole magnetic field which are initially twisted into direction of rotation (hereafter the positive twist). Rotation and plasma outflow in such a field strongly modify  the field line shapes. Nevertheless, apparently some fraction of the field lines remains positively twisted in the limits of the light cylinder in this case as well. These field lines are located  close to the field line separating closed and open field lines as it is shown in fig. \ref{fig0}. The shape of these field lines in the limits of the light cylinder strongly differs from the Archimedean spiral. This assumption does not contradict to the numerical simulations of the pulsar magnetosphere in force-free and MHD approximations \citep{iannis3,mnras2013}. Therefore, we can expect  that the plasma can be accelerated on these field lines due to the centrifugal mechanism. The objective of this paper is to consider how this mechanism operates  assuming that the positively twisted field lines  exist in the pulsar magnetosphere. Firstly, we intend to investigate how the energy of the plasma depends on the shape of the field line in 3-D space. 
For this we need an equation defining variation of the Lorentz factor of plasma along a magnetic field line of an arbitrary shape in 3-D.  
This equation is obtained in section 2. In section 3 we obtain an estimate of the maximal energy of the particles accelerated due to the 
magneto centrifugal mechanism of acceleration and discuss the results in sec. 4.

\section{Acceleration of plasma in crossed electric and magnetic fields}

\subsection{Variation of the Lorentz factor along flow lines}
 
We focus on the process of the plasma bulk motion  acceleration under the combined action of the electric and magnetic fields.  
Therefore it is reasonable to neglect thermal pressure to simplify the consideration.  Equation of motion in MHD approximation in this case takes the form
\begin{equation}
\rho \Omega \dot {\bf U}= q{\bf E}+{1\over c}[{\bf j}\times {\bf B}]  
\label{eq1}
\end{equation}
where $\bf E$ - electric field, $\bf j $ - density of the electric field, $\bf B$ - magnetic field, $\rho$ - density of plasma, $q$ - density of electric charge  and
the expression 
\begin{equation}
\dot {\bf U}={\partial {\bf U}\over \Omega \partial t} + v_k{\partial {\bf U}\over\partial x_k}.
\end{equation}
means full derivative of the four-velocity
$\bf U$ on time.  $\bf v$ is the velocity of plasma expressed in units of the light velocity $c$. Geometrical variables are measured in the units of the  radius of the light cylinder $R_c=c/\Omega$, where $\Omega$ is the angular  velocity of the pulsar rotation.
The rotational velocity in these variables is $V_{rot}=r$, where $r$ is the cylindrical radius.

 All the variables vary with time periodically.
We assume that the plasma flow satisfies to the frozen-in condition
\begin{equation}
{\bf E}+[\bf v\times \bf B]=0.
\label{frozen}
\end{equation}
This condition means that any particle of the plasma remains on the field line during the  motion. 

The Electric  field in the rotating magnetosphere is connected with the magnetic field as follows \citep{beskin1983}
\begin{equation}
 \bf E=-[V_{rot}\times B],
\label{efield}
\end{equation}
where ${\bf V}_{rot}=[{\bf e}_z\times \bf r]$.  This follows directly from the induction equation
\begin{equation}
{\partial {\bf B}\over \Omega \partial t}= -rot~ \bf E, 
\end{equation}
and the fact that the derivative of any vector field on time in the rotating magnetosphere can be presented as 
\begin{equation}
{\partial \bf B\over \Omega \partial t}= rot~[\bf V_{rot}\times \bf B].
\end{equation}

\begin{figure} 
\begin{center}
\includegraphics[width=64mm]{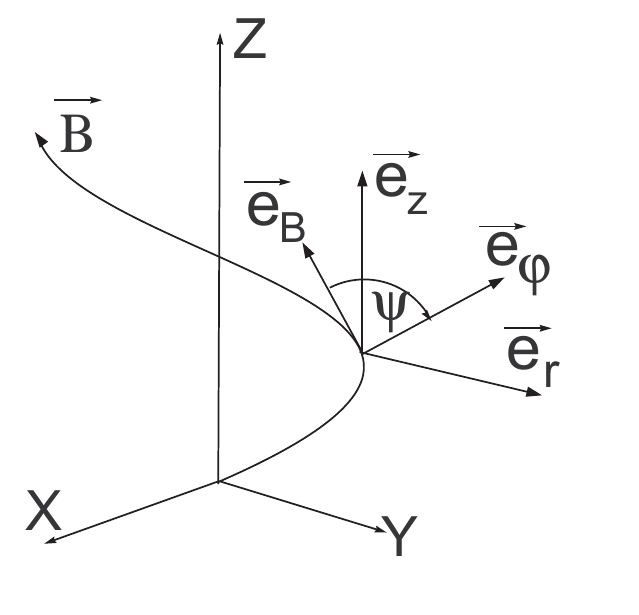}
 \caption{The geometry of the field lines and unit vectors.}
 \label{fig2}
\end{center}
\end{figure}

The velocity of plasma $\bf v$ can be presented as  a sum of velocity along the field line $\bf v_B$ 
and perpendicular to the field line $\bf v_d$.
The last one is the so called drift velocity defined from the frozen-in condition (\ref{frozen}) as follows
\begin{equation}
 \bf v_d=[\bf E\times \bf B]/B^2.
\label{vdrift}
\end{equation}
Let us  $\bf e_{\f}$,   $\bf e_r$ and $\bf e_z$  are the unit vectors of the cylindrical coordinate system along azimuthal, radial and axial directions. 
We introduce additional unit vectors $\bf e_d$, $\bf e_B$ along drift velocity and along the direction of the magnetic field line as shown in fig. (\ref{fig2}).
 Evidently, ${\bf e_d\cdot e_B}=0$.

 Substitution of (\ref{efield}) into (\ref{vdrift}) gives us 
\begin{equation}
 {\bf v}_d=V_{rot}(\bf e_{\f}-\bf e_B(\bf e_{\f}e_B)).
\label{drift}
\end{equation}

Equation for the Lorentz factor of plasma $\g$ can be obtained from (\ref{eq1}). Projection of this equation on the vector ${\bf U}_B$ gives
\begin{equation}
 {\bf U}_B \cdot  \dot {\bf U} = 0 ,
\label{first}
\end{equation}
where ${\bf U}_B=\g {\bf v}_B$. Taking into account the well known relativistic relationships $\g^2=1+U_d^2+U_b^2$ and $\g d\g= U_ddU_d+U_BdU_B$ and substituting into eq.
(\ref{first}) the equation ${\bf U}_B={\bf U}-{\bf U}_d$ we obtain
\begin{equation}
\g {\dot \g} = {\bf U}_d({\dot {\bf U}_d}+{\dot {\bf U}_B}). 
\label{second}
\end{equation}
The product $\bf U_d\cdot \bf U_B=0$. Therefore
\begin{equation}
  {\bf U}_d \dot {\bf U}_B=-{\bf U}_B{\dot {\bf U}_d}
\end{equation}
and eq. (\ref{second}) takes a form
\begin{equation}
 \g {\dot \g} = ({\bf U}_d-{\bf U_B}){\dot {\bf U}_d}
\end{equation}
After substitution into this equation the time derivative ${\dot {\bf U}_d}=\g {\dot {\bf v}_d}+\bf v_d{\dot \g}$ we obtain the following equation
\begin{equation}
 (1-v_d^2){\dot \g}=\g (\bf v_d-\bf v_b) {\dot {\bf v}_d}.
\label{basic}
\end{equation}
It is convenient to consider two parts of the right hand part of this equation separately.
The first part takes a form
\begin{equation}
{ \bf v}_d{\dot  {\bf v}_d}={1\over 2}\dot {v_d^2}.
\end{equation}
$v_d^2$ is defined from (\ref{drift}) as $v_d^2=V_{rot}^2(1-\cos{\psi}^2)$, where $\psi$ is the angle between $\bf e_{\f}$ and $\bf e_B$ and 
$\cos{\psi}= \bf e_{\f}\cdot\bf e_B $.  Therefore this part of the equation takes a form
\begin{equation}
 { \bf v}_d{\dot {\bf v}_d}=V_{rot}\sin{\psi}^2\dot V_{rot}-V_{rot}^2\cos{\psi} \dot{ cos{\psi}}.
\label{eq15}
\end{equation}

Taking into account eq. (\ref{drift}) it is easy to obtain that
\begin{equation}
 {\dot {\bf v}_d}=\dot V_{rot}(\bf e_{\f}-\bf e_B\cos{\psi})+V_{rot}(\dot {\bf e}_{\f}- \bf {\dot e}_{B} \cos{\psi}-\bf e_B\dot{\cos{\psi}})
\label{eq16}
\end{equation}

Along the trajectory of the particles 
\begin{equation}
 \dot {\bf e_{\f}} = -v_{\f}{{\bf e}_r\over r},
\label{eq17}
\end{equation}
and ${\bf e}_B{\bf \dot e}_{B}=0$, because ${\bf e}_B\cdot {\bf e}_B=1$. Taking both conditions into account we obtain that 
\begin{equation}
 {\bf v}_B {\dot {\bf v}_d}=-V_{rot}v_B({v_{\f}({\bf e}_r{\bf e}_B)\over r}+\dot{\cos{\psi}}).
\label{eq18} 
\end{equation}
Substitution of eqs. (\ref{eq15}) and (\ref{eq18}) into eq. (\ref{basic}) gives that
\begin{eqnarray}
 (1-v^2_d)\dot \g=\g(r v_r\sin{\psi}^2+v_B v_{\f}({\bf e}_r{\bf e}_B)+    && \\ \nonumber
 +r(v_{B}-V_{rot}\cos{\psi})\dot{\cos{\psi}}). & & 
\end{eqnarray}
Here we took into account that $V_{rot}=r$ and $\dot V_{rot}=v_r$.

This equation can be simplified. The  $r$ component of the full velocity $\bf v= v_d+ v_B$ can be presented using eq. (\ref{drift}) as 
\begin{equation}
  v_r=({\bf e}_r{\bf e}_B)(v_B-V_{rot}\cos{\psi}),
\label{eq20}
\end{equation}
while  $\f$ component of the velocity equals to
\begin{equation}
v_{\f}=v_B\cos{\psi}+V_{rot}\sin^2{\psi}
\label{eq20a} 
\end{equation}
Then 
\begin{equation}
 (1-v^2_d)\dot \g=\g(rv_r\sin{\psi}^2+v_B v_{\f}({\bf e}_r{\bf e}_B)+{r v_r\dot{\cos{\psi}}\over({\bf e}_r{\bf e}_B)} ).
\label{eq22}
\end{equation}

\subsection{Variation of the Lorentz factor along magnetic field lines}

Equation (\ref{eq22}) defines the variation of the Lorentz factor of plasma along the trajectory of the particles. Nevertheless, they remain
 on the rotating field line at the motion due to frozen-in condition.  This allows us to obtain an equation defining 
the variation of the Lorentz factor of the particles along any fixed field line. Let's  pay attention that any scalar satisfies to 
the equation
\begin{equation}
 \g (t+\delta t,{\bf r})=  \g (t,{\bf r}-{\bf V}_{rot}\delta t)
\end{equation}
at the stationary rotation of the magnetosphere. Therefore the full time derivative of $\g$ along the plasma trajectory transforms into 
\begin{equation}
 \dot \g= (\bf v-\bf V_{rot}){\partial \g\over \partial \bf r}.
\end{equation}
After substitution into this equation of the relationship ${\bf v- V}_{rot}={\bf e}_B(v_B-V_{rot}\cos{\psi})$ we obtain   
\begin{equation}
  \dot \g=v_r {{\bf e}_B \partial \g\over ({\bf e}_r{\bf e}_B)\partial \bf r}
\end{equation}
The derivative ${\bf e}_B {\partial \g\over \partial\bf r}$ is taken  along the filed line. The full derivative 
of the Lorentz factor on $t$ is transformed into  derivative of the Lorentz factor on the cylindrical radius $r$ taken along the field line as follows
 \begin{equation}
  \dot \g=v_r {\partial \g\over \partial r}
\end{equation}
For the equation defining the variation of the Lorenz factor along a field line in 3-D space we obtain
\begin{eqnarray}\label{eq26}
 (1-v^2_d)v_r {\partial \g\over \partial r}=\g(r v_r\sin^2{\psi}+v_B v_{\f}({\bf e}_r{\bf e}_B)+&& \\ \nonumber
+{r v_r\dot{\cos{\psi}}\over({\bf e}_r{\bf e}_B)} ), &&
\end{eqnarray}
where 
\begin{equation}
 \dot{\cos{\psi}}=v_r{\partial \cos{\psi}\over \partial r}.
\end{equation}

\section{General properties of the flow}

\subsection{Crossing the light cylinder}
%Plasma crosses the light cylinder only if $\cos{\psi} < 0$. 
Radial velocity of  plasma is defined by eq. (\ref{eq20}). Simultaneous multiply and division of this equation on  $ (v_B+V_{rot}\cos{\psi})$ gives 
\begin{equation}
 v_r=({\bf e}_r{\bf e}_B){(v^2-V_{rot}^2)\over (v_B+V_{rot}\cos{\psi})}.
\label{eq27}
\end{equation}
The nominator of this equation goes to zero at $r=v$.  This point is located very close to the light cylinder for the relativistic plasma.  Radial velocity of plasma goes to zero in this point provided that $\cos{\psi} > 0$ (positive twist). Plasma can not cross this point. In the opposite case  $\cos{\psi} < 0$ the denominator of eq. (\ref{eq27}) can go to zero in this point simultaneously with the nominator remaining $v_r$ nonzero. Thus, the plasma can cross this point only if the field line has negative twist at $r=v$.    
Physical sense of this condition is rather transparent. Below we will see that eq. (\ref{eq26}) gives infinite Lorentz factor in the point $r=v$ provided that $\cos{\psi} > 0$. This means that the inertia of plasma infinitely increases at approaching to the  point $r=v$ and eventually the inertia of plasma  at some moment changes the twist of the magnetic field on the negative.  This way in the self consistent solutions the twist of the magnetic field lines is negative at the point $r=v$ always. Therefore, below we should consider only physically admissible field lines having negative twist at $r=v$.

\subsection{The Archimedean spiral is an attractor}

The shape of the magnetic field line in 3-D can be parametrised by many ways. In this paper we assume that the cylindrical radius $r$ monotonically varies 
at the motion along the  field line from the surface of the star. This is apparently valid for all open field lines along which plasma flows from the magnetosphere. In this case the field line in 3-D can be parametrised by functions $\Phi(r)=\f-\Omega t$, azimuthal angle of the field line and $Z(r)$, z coordinate of the field line. They do not depend on time because every field line simply rotates with constant angular velocity.   In this case the unit vector ${\bf e}_B$ can be presented as 
\begin{equation}
 {\bf e}_B = {({\bf e}_r+{r\partial \Phi\over \partial r}{\bf e}_{\f}+{\partial Z\over \partial r}{\bf e}_z)\over \sqrt{1+({r\partial \Phi\over \partial r})^2+({\partial Z\over \partial r})^2}}.
\end{equation}
The magnetic field line  frozen into plasma forms an Archimedean spiral if the plasma flows radially with constant velocity $V_0$.  In this case $\Phi(r)$  depends on $r$ as 
\begin{equation}
 \Phi=-{r\over \sin{\alpha}V_0},
\end{equation}
where $\alpha$ is the polar angle of the plasma velocity. Function $Z(r)$ depends on $r$ as follows 
\begin{equation}
 Z=r\cot{\alpha}. 
\end{equation}
Therefore 
\begin{equation}
 ({\bf e}_r{\bf e}_B)= {\sin{\alpha}\over \sqrt{1+({r\over V_0})^2}},
\end{equation}
and
\begin{equation}
 \cos{\psi}=({\bf e}_r{\bf e}_{\f})= -{r\over V_0\sqrt{1+({r\over V_0})^2}},
\end{equation}
while 
\begin{equation}
{\partial \cos{\psi}\over \partial r}= -{1\over V_0}{1\over (1+({r\over V_0})^2)^{{3\over 2}}}.
\end{equation}
It is convenient to present this derivative through the products  $({\bf e}_B{\bf e}_{\f})$ and  $({\bf e}_B{\bf e}_{r})$. In this case  
\begin{equation}
 {\partial \cos{\psi}\over \partial r}={1\over r} \cos{\psi}\sin{\psi} ({{\bf e}_B{\bf e}_{r}\over \sin{\alpha}}).
\end{equation}
It is easy to show that plasma with the  velocity $V_0$ has $v_B=V_0\sin{\psi}$, $v_r=V_0\sin{\alpha}$ and $v_{\f}=0$. 
Substitution of these equations into eq. (\ref{eq26}) gives ${\partial \g\over \partial r}=0$. Thus, the plasma flows in this magnetic field with the 
constant Lorentz factor. The Lorentz factor $\g_0={1\over \sqrt{1-V_0^2}}$ is stationary point of eq. (\ref{eq26}). The deviation $\delta\g$ of the Lorentz factor 
is defined by the equation
\begin{equation}
(1-v^2_d) v_r{\partial \delta\g\over \partial r}= 2\g_0 \delta v_{\f}V_0\sin^2{\psi}\sin{\alpha}.
\label{eq36}
\end{equation}
It follows from (\ref{eq20a}) that 
\begin{equation}
  \delta v_{\f}=\delta v_B\cos{\psi}
\end{equation}
while 
\begin{equation}
 \delta v_B={\delta\g\over v_B\g_0^3}.
\end{equation}
In the primary flow $v_B=V_0\sin{\psi}$, $\sin{\psi}\cos{\psi}=-{r\over V_0(1+({r\over V_0})^2)}$ and 
\begin{equation}
 1-v_d^2={1+({r\over U_0})^2\over 1+({r\over V_0})^2}.
\end{equation}
At  distance $r \ll U_0$ the equation for $\delta \g$ takes a form
\begin{equation}
 {\partial \delta\g\over \partial x}=-{\delta\g\sin{\alpha}\over \g_0^2},
\end{equation}
where $x=(r/V_0)^2$.

\begin{figure} 
\begin{center}
\includegraphics[width=84mm]{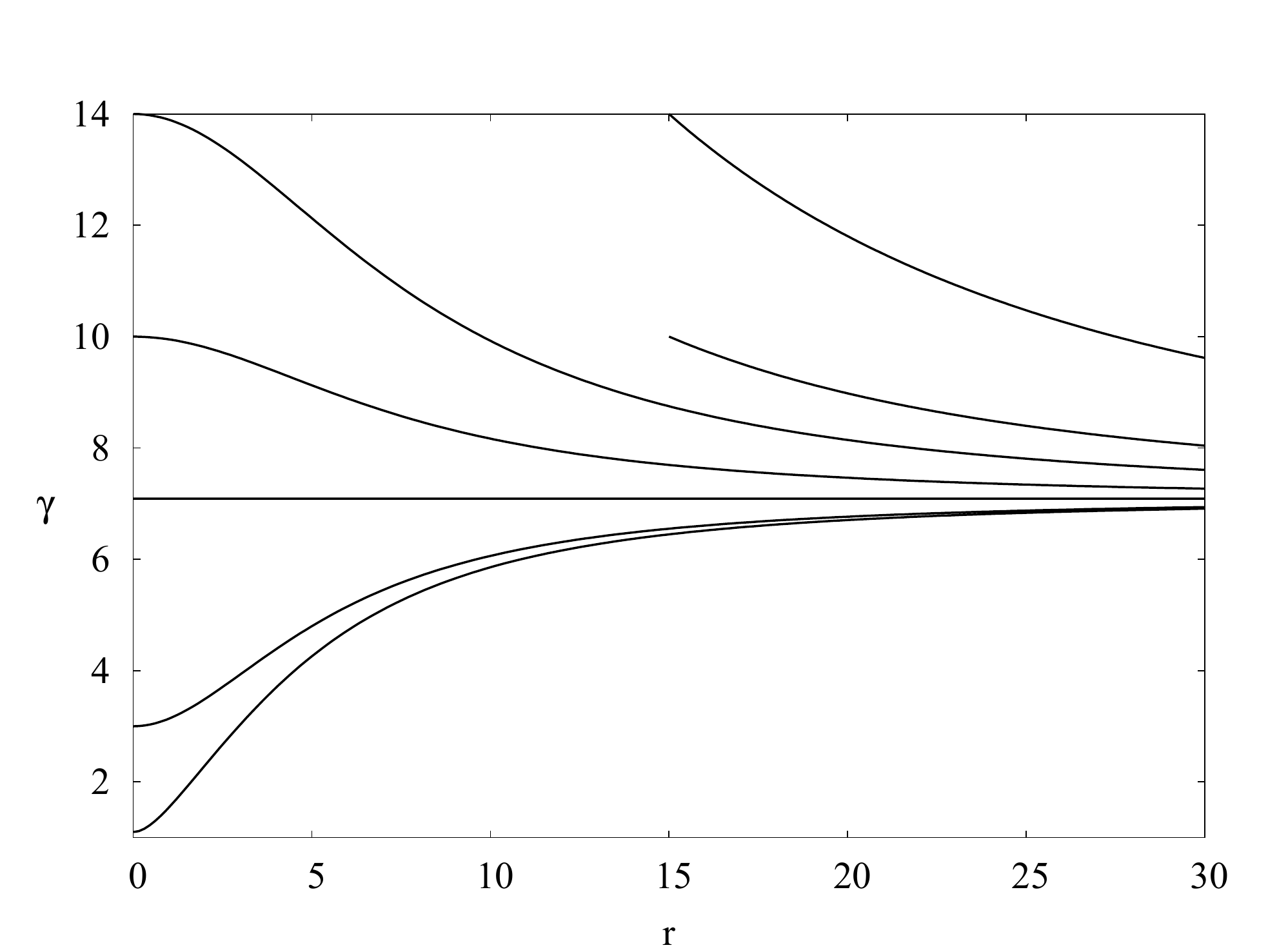}
 \caption{Dependence of Lorentz factor on radius in the magnetic field of Archimedean spiral with $V_0=0.99$ and polar angle $\alpha=90^0$. The horizontal 
line shows the Lorentz factor for $\g=\g_0$.  One family of lines starts from $r_0=0.01$, another family starts at $r_0=15$. All the trajectories converge to 
$\g=\g_0$.}
 \label{fig3}
\end{center}
\end{figure}

It follows from this equation that any deviation   $\delta\g$ goes to zero with $r$. This means that the trajectory $\g(r)=\g_0$ attracts all other 
trajectories.
To understand the behavior of the initial deviation of the Lorentz factor at the conditions when $\delta\g \approx \g_0$  eq. (\ref{eq26}) has been solved 
numerically. The result is shown in fig.  \ref{fig3}.

The solution confirms that the stationary point $\g=\g_0$ is the attractor even for large deviation of $\g$ from $\g_0$.

\subsubsection{Acceleration at $V_0 \ge 1$}

\begin{figure} 
\begin{center}
\includegraphics[width=84mm]{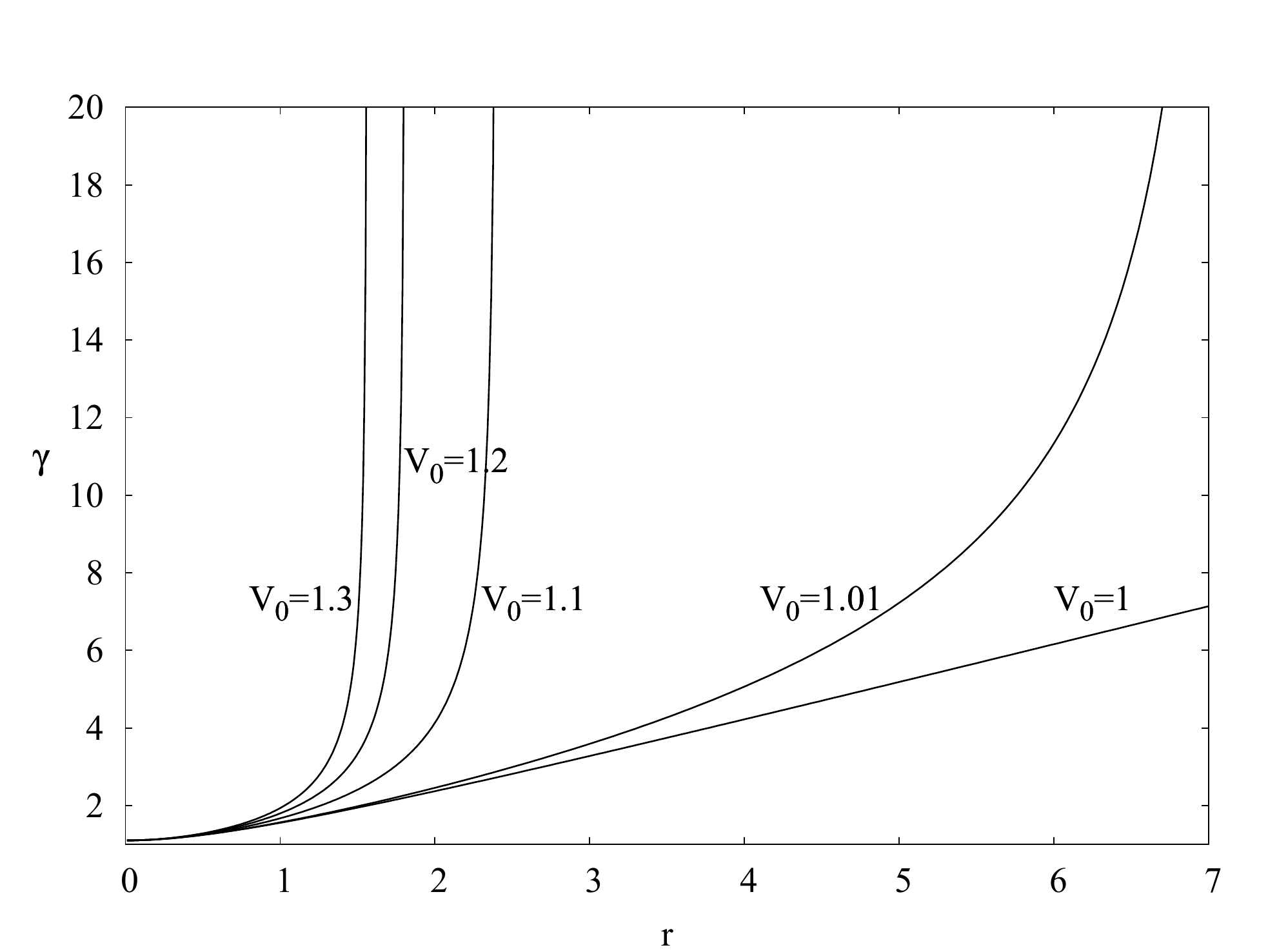}
 \caption{Dependence of the Lorentz factor of plasma on $r$ for different parameters $V_0 \ge 1$}
 \label{fig4}
\end{center}
\end{figure}

The case  $V_0 > 1$ looks unphysical at first glance.  In this case $V_0$ should be considered as a parameter defining the twist of the spiral, not the actual
velocity of the plasma forming the magnetic field. 
We consider this example to understand the behaviour of plasma when the field lines are twisted 
less than it happens in the limiting case of the radial wind expanding with a constant velocity.  Fig. \ref{fig4} shows the dependence of the Lorentz factor on r for a range of $V_0 \ge 1$. The plasma is accelerated linearly with $r$ if  $V_0=1$.  It is possible to make sure by direct substitution into eq. (\ref{eq26}) that the solution $u=r$ is the exact solution of eq.(\ref{eq26}) for the perfect Archimedean spiral with the parameter $V_0=1$. This corresponds to the behaviour of the Lorenz factor in the force-free model  with the  split monopole magnetic field \citep{narayan}.  

At $V_0 > 1$ the Lorentz factor diverges at final distance from the center. The position where this happens is defined by the condition $v_d=1$. In framework of our parametrization of the field line this condition is fulfilled at distance $r={V_0\over \sqrt{V_0^2-1}}$. According to this equation the point where the plasma gets infinite Lorentz factor is located at a few light cylinders even at a small deviation of $V_0$ above 1. Fig. \ref{fig4} shows that $V_0=1.01$ results to divergence of the Lorentz factor at distance $r \sim 7$ light cylinders. 
 Here we firstly deal with the divergence of the Lorentz factor in some point. It is necessary to keep in mind that this divergence appears only because we use the prescribed shape of the magnetic field line. In real self consistent solution the growth of the Lorentz factor results into growth of the inertia of the plasma which obviously  increases the twist the magnetic field preventing infinite growth of the plasma Lorentz factor. Therefore, the terminal energy of the plasma can be defined only in the self consistent solution. Nevertheless, our approach allows us to specify the most interesting regions where the efficient acceleration should take place. 

 This particular example of motion in the field of perfect Archimedian spiral allows us to make rather general conclusion. If the negative twist is less than the twist of the Archimedian spiral with $V_0=1$, then the plasma can be essentially accelerated down stream the light cylinder. Obviously,  this behaviour  was found by  \cite{mnras2013}. The distribution of the four-velocity in the equatorial plane from their work is shown in fig \ref{Chehov}. We selected for example  two field lines numbered as 1 and 2.  Field line 1 was twisted negatively in the original dipole magnetic field. There is only weak acceleration of the plasma along this line. Field 2 was twisted positively in the original vacuum dipole magnetic field. Plasma is accelerated rather efficiently down stream the light cylinder along this field line. The acceleration occurs faster than $u=r$ (see fig. 2c from \citep{mnras2013}). This means that field line 2 is twisted in the negative direction less than the Archimedian spiral with  $V_0=1$. This example shows that the initially positively twisted field lines ( in the sense of the initial dipole magnetic field) can provide remarkable acceleration of plasma even down stream the light cylinder,  exactly in accordance with the predictions of the work \citep{bog2001}. Of course, only full self consistent solution of the problem in MHD approximation can define the terminal Lorentz factor of the plasma in this case.

\subsection{Magnetocentrifugal acceleration at $V_0 < 0$.}

%{\bf Acceleration of plasma in the limits of the light cylinder by the rotating magnetic field we call here as the %magneto centrifugal acceleration.

  Below we show that the law of the 
Lorentz factor variation along the field lines is rather universal for the field lines having positive twist and estimate the possible maximal Lorentz factor which can be achieved due to this mechanism.

\subsubsection{Solution of the equation}
It is convenient to consider equation (\ref{eq26}) in the ultrarelativistic limit. This approximation is reasonable because even the initial Lorentz  factor of plasma in the pulsar magnetosphere exceeds 10-100.  In this limit 
$v=1$ and the component of the velocity  along the magnetic field equals to
\begin{equation}
 v_B=\sqrt{1-v_d^2}.
\label{eq37}
\end{equation}
%Let us take into account that
%\begin{eqnarray}
% V_{rot}v_r\sin{\psi}^2+v_B v_{\f}({\bf e}_r{\bf e}_B)=V_{rot}v_r\sin{\psi}^2+ & & \\ \nonumber
%({\bf e}_r{\bf e}_B)v_B(v_B\cos{\psi}+v_{rot}\sin{\psi}^2). &&
%\label{eq38}
%\end{eqnarray}
Using eq. (\ref{eq20}) and (\ref{eq20a}) it is easy to obtain that 
\begin{eqnarray}
 V_{rot}v_r\sin{\psi}^2+v_B v_{\f}({\bf e}_r{\bf e}_B)=({\bf e}_r{\bf e}_B)(2v_BV_{rot}\sin{\psi}^2+ && \\ \nonumber
+(2v_B^2-v^2)\cos{\psi}). & &
\end{eqnarray}
Let us to divide this equation on $v_r$ taking into account that this variable is defined by eq. (\ref{eq27}).  Simple transformations give
\begin{eqnarray}
 (V_{rot}v_r\sin{\psi}^2+v_B v_{\f}({\bf e}_r{\bf e}_B))\times ({v_B+V_{rot}\cos{\psi}\over {\bf e}_r{\bf e}_B})= &&\\ \nonumber
=(2v_B^2V_{rot}+{v^2v_r\cos{\psi}\over ({\bf e}_r{\bf e}_B)}) &&
\end{eqnarray}
Taking this into account we obtain
\begin{equation}
 {\partial \g\over \g\partial r}={2r\over (1-r^2)}+{v_r(v^2\cos{\psi}+ r(1-r^2){\partial {\cos{\psi}}\over \partial r})\over (1-r^2)({\bf e}_r{\bf e}_B)(1-v_d^2)}.
\label{eq42}
\end{equation}
We used here also eq. (\ref{eq37}) and the fact that $V_{rot}=r$.

At the light cylinder $r = 1$ the first term in the right hand part of eq. (\ref{eq42}) diverges while the second one remains finite. Therefore, the variation  of the 
Lorentz factor at the light cylinder is defined basically by the approximate equation
\begin{equation}
  {\partial \g\over \g\partial r}={2r\over (1-r^2)},
\end{equation}
which has  trivial solution 
\begin{equation}
 \g={\g_0 (1-r_0^2)\over (1-r^2)}.
\label{eq48}
\label{gamma}
\end{equation}
This solution does not depend on the shape of the field line provided that the field line is positively  twisted. In this case 
$\cos{\psi} \ge 0$ and this provides us that the second term in the right hand part of eq. (\ref{eq42}) is small compared  with the first one at the light cylinder.
\begin{figure} 
\begin{center}
\includegraphics[width=84mm]{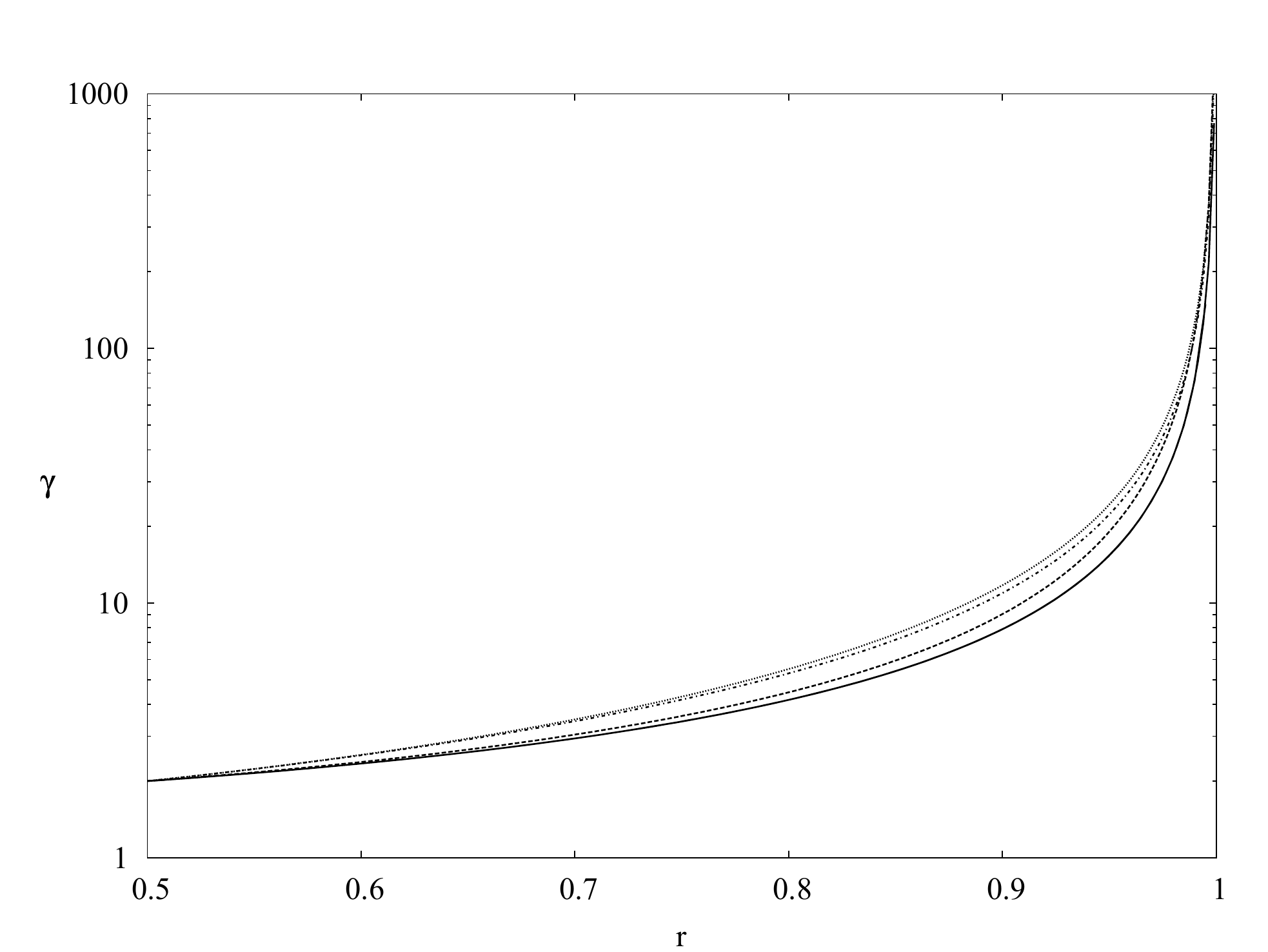}
\caption{Dependence of Lorentz factor on $r$. Lower curve is solution (\ref{gamma}). Dashed, dashed-doted and dotted curves are the solution of eq. (\ref{eq27}) for $V_0=-10, -1$ and $-0.1$ correspondingly. Lower solid line - solution(\ref{eq48}}
 \label{fig5}
\end{center}
\end{figure}
It is easy to obtain that exactly the same law of variation can be obtained for the Lorentz factor of a particle moving along  a rotating straight solid wire. This is the 
law of acceleration due to the magneto centrifugal force.  This dependence for the acceleration on a straight field line  directly follows from eq. (\ref{eq42}) if to take into account that for these field lines $\cos{\psi}=0$ and ${{\partial \cos{\psi}}\over \partial r}=0$. 

To make sure  that the behaviour of the Lorentz  factor at the light cylinder does not depend on the shape of the field line eq. (\ref{eq26}) has been solved for the 
field lines having shape of the Archimedean spiral but with $V_0 < 0$. The result is shown in fig. \ref{fig5} in comparison with solution (\ref{gamma}). It follows from this figure that  all the field lines give universal solution for the Lorentz factor at the light cylinder. 

\begin{figure}
\includegraphics[width=84mm]{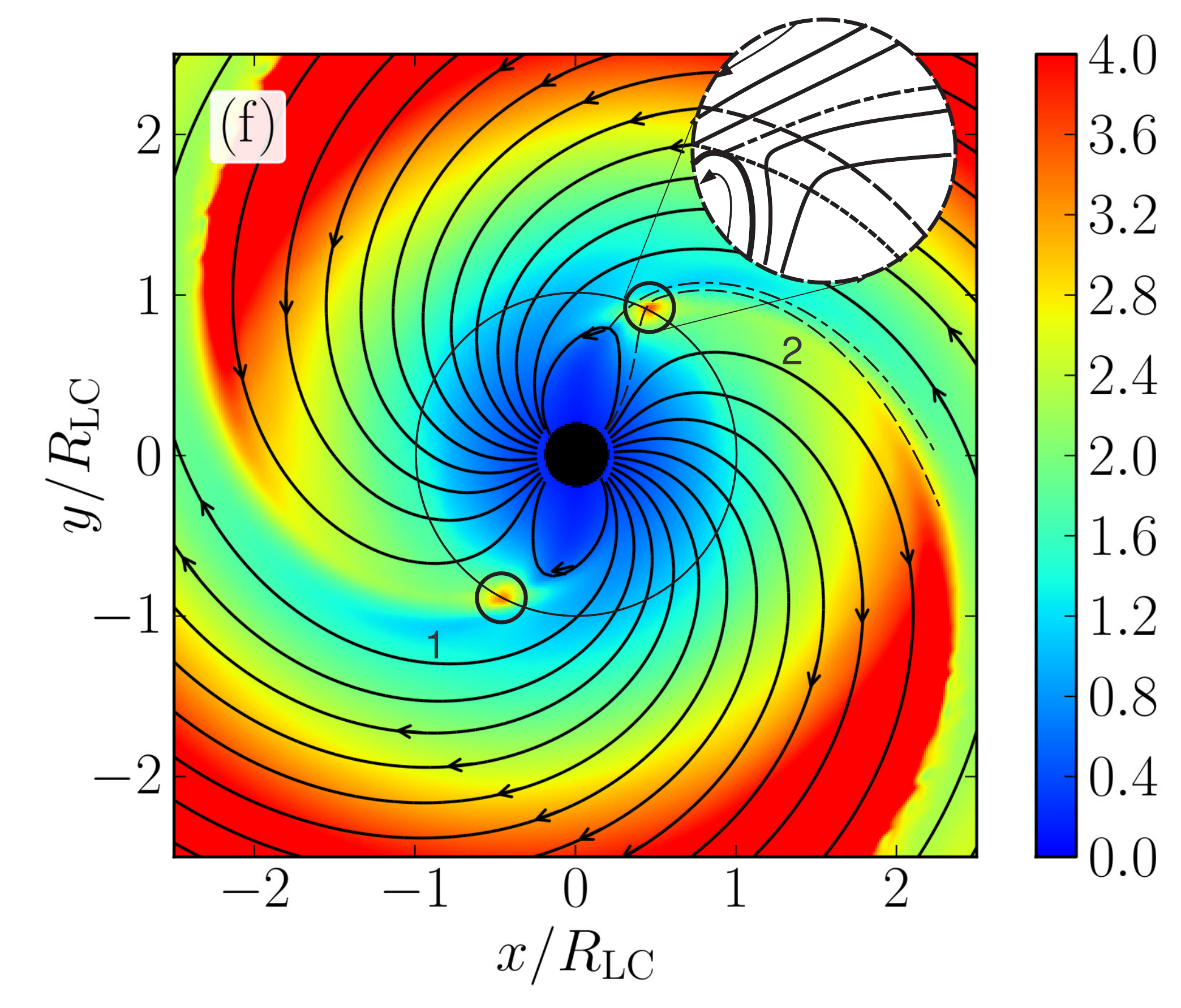}
 \caption{Distribution of the plasma four-velocity in the equatorial plane of a pulsar rotating under angle $60^0$ taken from \citep{mnras2013}. Solid circles point onto locations of remarkable acceleration of plasma at the light cylinder presumably due to centrifugal mechanism. Dashed lines show possible shape of the field lines crossing the light cylinder in this place. Inclusion at the upper right corner shows in more details the structure of the flow and magnetic field at the place of acceleration. 
Dashed line - light cylinder, dotted line - Alfven surface, Dashed-dotted line - separatrix field line.}
 \label{Chehov}
\end{figure}

 It is interesting again to compare our results with the results of direct calculations of the pulsar magnetosphere in MHD approximation.  Fig. \ref{Chehov} is taken from \cite{mnras2013}. This figure shows distribution of the four-velocity in the equatorial plane of the pulsar having the angle of inclination $60^0$. There is interesting evidence of the plasma acceleration at the light cylinder close to the separatrix field line ( or current sheet). This region is shown  by circles.  Unfortunately, the shape of the field lines in the most interesting region is not shown by the authors in this figure. Therefore, we draw
our assumptions about the shape of the field line in this region by dashed lines. 
The circled inclusion at the right upper corner of the figure shows the possible structure of the field lines close to the region at the light cylinder when the closed field lines reach the Alfvenic surface.     

\subsection{Maximal energy of electrons}

The Lorentz factor of plasma diverges and  the radial velocity goes to zero at the light cylinder if $\cos{\psi} > 0$. The passage through the 
light cylinder is impossible if the field line has positive twist.  All open field lines must have negative twist here. Therefore, the largest Lorentz factor which can be achieved by plasma is defined by the  location where the field lines change the direction of twist. According to the geometry presented in fig. \ref{fig0} the field lines closest to the last closed field line have  positive twist up to 
the point where the current sheet separating the magnetic fluxes of opposite polarities takes start. The question about structure of the magnetic field in  this region is not investigated in ideal MHD approximation. In force-free approximation the closed magnetic field lines can reach the light cylinder \citep{gruzinov,timokhin,iannis1}. However this is not true in the case of MHD flow \citep{iannis2}. At present it is possible only to speculate about the largest distance from the star where the closed filed lines can reach because even a direct calculation of the structure of the magnetosphere near the light cylinder does not give 
unambiguous answer \citep{komissarov}.  In this paper we  assume that the closed field lines can reach the Alfvenic surface where the velocity of plasma equals to the local Afvenic velocity as it is shown in inclusion into fig. \ref{Chehov}.

The velocity of plasma equals to the local Alfvenic velocity in the Alfvenic point located at distance $r_A$ from the star. In the frame system where the electric field equals to zero this condition can be expressed by the following equation
\begin{equation}
 (1-{B_c^2\over 4\pi \rho^* U_c^2})=0.
\label{eq45}
\end{equation}
Here $U_c$ and $B_c$ are four velocity and magnetic field in the system coordinate comoving  with the velocity of rotation at the Alfvenic point. $\rho^*$  is the density of plasma in the comoving coordinate system where the plasma velocity equals to zero. In the system comoving with the rotational velocity at the Afvenic surface the ratio
${B_c^2\over U_c^2}={B_{rc}^2\over U_{rc}^2}$, where $B_{rc}$ and $U_{rc}$ are the radial components of the magnetic field and four-velocity in the comoving system. Now let us to use  that the magnetic field $B_{rc}$ is connected with the magnetic field in the laboratory frame system as follows
\begin{equation}
 B_{rc}^2=B_r^2-E^2_z=(1-r^2_A)B_r^2.
\end{equation}
This is the consequence of the relativistic invariant $B_r^2-E^2_z=const$. The component of the four velocity $U_r$ is the relativistic invariant as well. Therefore $U_r=U_{rc}$. Taking all this into account eq. (\ref{eq45}) can be rewritten as 
\begin{equation}
(U_r-{(1-r^2_A)B_r^2\over 4\pi \rho^* c^2 U_r})=0.
\label{eq47}
\end{equation}

The continuity equation 
\begin{equation}
 {\partial \rho\over \partial t}+{\partial \rho {\bf v}\over \partial r}=0 
\end{equation}
can be rewritten for the steady state rotation of the magnetosphere in the form ${\partial \rho ({\bf v}-{\bf V_{rot}})\over \partial r}=0$. Taking onto account that
${\bf v-V}_{rot}={\bf e}_B{v_r\over ({\bf e}_r{\bf e}_B)}$ the last equation can be presented in the form
\begin{equation}
 {\partial\over \partial r} {{\bf B}\rho v_r\over ({\bf e}_r{\bf e}_B)B}={\partial\over \partial {\bf r}} {\bf B}{\rho v_r\over B_r}=0.
\end{equation}
This means that the ratio $ {\rho v_r\over B_r}$ is constant along the field line. Therefore, eq. (\ref{eq47}) can be rewritten as 
\begin{equation}
 (U_r-(1-r^2_A)B_r {B_{r0}\over \rho_0 c^2 v_{r0}})=0,
\label{eq50}
\end{equation}
where $B_{r0}$, $v_{r0}$ $\rho_0$ are the initial radial components of the magnetic field and velocity as well as initial density of plasma.

Dependence on $r$ of the velocity near the light cylinder can be obtained from eq. (\ref{eq27}) as 
\begin{equation}
 v_r=({\bf e}_r{\bf e}_B) {(1-r^2_A)\over 2\cos{\psi}}.
\end{equation}
This means that 
\begin{equation}
 U_r=({\bf e}_r{\bf e}_B){\g_0\over 2\cos{\psi}}
\end{equation}
remains finite at the light cylinder.
Substitution of the last equation into eq. (\ref{eq50}) gives that
\begin{equation}
 (1-r^2_A)={4\pi\rho_0 c^2({\bf e}_r{\bf e}_B)\g_0 \over 2\cos{\psi}B_0^2}({R_0\over R_L})^3,
\label{distance}
\end{equation}
where we assumed that the magnetic field at the light cylinder is connected with the field $B_0$ at the surface of the neutron star as $B_r \approx B_0({R_0\over r_L})^3$, where $R_0$ is the radius of the star.  This equation allows us to estimate the relative difference between the radius of the Alfvenic point and the light cylinder. It equals to 
\begin{eqnarray}
 {\delta r\over R_L}={2\pi \rho_{GJ} \g_G mc^2({\bf e}_r{\bf e}_B)\over B^2({\bf e}_B{\bf e}_{\f})}({R_0\over R_L})^3= &&\\ \nonumber
=10^{-4}({\g_G\over 10^7})({33 ~ms\over P})^4({4.4 \cdot 10^{12} {\rm G}\over B})({R_0\over 10 {\rm km}})^3({{\bf e}_{r}{\bf e}_B\over {\bf e}_{\f}{\bf e}_B})
\end{eqnarray}
Here $\rho_{GJ} ={\Omega B\over 2\pi e c}$ is the Goldreich-Julian density of particles \citep{gj} and $\g_G \sim 10^7$ is the Lorentz factor to which an electron is accelerated in the electrostatic gaps.

Substitution of eq (\ref{distance}) into  eq. (\ref{gamma}) gives an estimate of the maximal energy of the accelerated electrons as follows
\begin{equation}
 \g_{max}={2({\bf e}_{\f}{\bf e}_B)\over ({\bf e}_r{\bf e}_B)}{B_0^2R_0^3\over 4\pi\Lambda\rho_{GJ}c^2},
\end{equation}
where $\Lambda\sim 10^3$ is the multiplication factor showing the effect of the multiplication of the plasma in the electromagnetic cascade in the magnetosphere \citep{harding}.
Estimates give for $\g_{max}$
\begin{equation}
\g_{max}=5\cdot 10^7 ({10^3\over \Lambda}) ({R_0\over 10 {\rm km}})^3 ({33 ms\over P})^2 ({B\over 4\cdot 10^{12} \rm G})({{\bf e}_{\f}{\bf e}_B\over {\bf e}_r{\bf e}_B}).  
\end{equation}
It follows from these estimates that for Crab parameters the Lorentz factor can  achieve value $5\cdot 10^7$ which is quite sufficient to  explain the observed pulsed radiation from this pulsar.

\section{Discussion}

We show that in ideal MHD approximation the energy of the plasma is defined entirely by the 3-D geometrical shape of the magnetic field lines. 
Eq. (\ref{eq26}) can be used for a posteriori calculation of the energy of plasma provided that the shape of the field lines are calculated some way.  
This can be done for example in the case of  calculation of the pulsar magnetosphere in force-free approximation performed in \cite{spitkovskii,iannis3}. Energy of the plasma remains undefined in this approximation.   

The most interesting result is the possibility of magneto centrifugal acceleration of plasma bulk motion close to the light cylinder. This mechanism can provide acceleration of plasma up to $\g \sim 10^7$ for the Crab parameters provided that the region of closed field lines achieve the Alfvenic surface. In this case a positively twisted flux of magnetic field lines  stretched up to the Alfvenic surface exists. Currently it is impossible to predict the amount of the magnetic and particle flux of these field lines. This problem should be specially investigated in numerical calculations of the pulsar magnetosphere. 

It follows from our consideration that the magnetocentrifugal mechanism can provide acceleration only of a small fraction of the electrons of the wind. All the rest of the wind will move with the bulk Lorentz factor of the order of $\g_0$. In this regard it becomes interesting the interaction of the energetic electrons with the electrons of the wind. However, consideration of this problem is beyond of scopes of the present paper. 

The existence of the field lines where the magneto centrifugal acceleration  can occur opens new way for interpretation of the observed pulsed VHE radiation from the  Crab pulsar with energy $\sim 400 ~ \rm GeV$. In the work  \citep{aharonian} this radiation has been interpreted as the inverse Compton radiation from the electrons of the wind accelerated $\sim 30$  light cylinders downstream the light cylinder provided that all the electrons of the wind are accelerated to the Lorentz factor $\sim 10^6$.  Actually, our analysis combined with the results of \citep{mnras2013} confirms that the efficient acceleration not all but remarkable fraction of the wind electrons can occur beyond the light cylinder. Nevertheless, the situation dramatically changes if only a tiny fraction of the electrons is accelerated at the light cylinder. In this case there is no overproduction of VHE gamma-rays and the observed pulsed VHE gamma-rays can be generated at the light cylinder. If so, we immediately come to the conclusion that  all the rest of the radiation from the Crab pulsar is generated at the light cylinder in the same place where the pulsed VHE gamma-rays are generated because all the radiation is well synchronized in phase. 
This is natural from physical point of view. The twist of the magnetic field should vary very fast at the place of acceleration. Remind that the twist should change from positive at the Alfvenic surface to negative at the light cylinder. Therefore, it is naturally to expect generation of rather intensive curvature radiation from the same electrons. All this shows that the region close to the last closed field line can be very interesting for the numerical investigation of the physics of pulsars in ideal MHD approximation.

\section*{Acknowledgments}
 The author is grateful to Dr. F.Aharonian for useful discussions and unknown reviewer for interesting comments.

\label{lastpage}

\end{document}